\documentclass[aps,nofootinbib]{revtex4}
\usepackage{graphicx}
\usepackage{amssymb}

\begin{document}

%============================================================================
\title{Probing singularities in quantum cosmology with curvature 
scalars}

\author{G. Oliveira-Neto\footnote{Corresponding author: gilneto@fat.uerj.br},
E. V. Corr\^{e}a Silva}
\affiliation{Departamento de Matem\'{a}tica e Computa\c{c}\~{a}o, 
  	Faculdade de Tecnologia, Universidade do Estado do Rio de Janeiro \\
 	Rodovia Presidente Dutra, km 298, P\'{o}lo Industrial, 
 	CEP 27537-000, Resende, RJ, Brasil.}

\author{N.A. Lemos}
\affiliation{Instituto de F\'{\i}sica,  
           Universidade Federal Fluminense, 
           R. Gal. Milton Tavares de Souza s/n,
           CEP 24210-340 - Niter\'{o}i, RJ - Brasil.}

\author{G.A. Monerat}
\affiliation{Departamento de Matem\'{a}tica e Computa\c{c}\~{a}o, 
  	Faculdade de Tecnologia, Universidade do Estado do Rio de Janeiro \\
 	Rodovia Presidente Dutra, km 298, P\'{o}lo Industrial, 
 	CEP 27537-000, Resende, RJ, \ Brasil.}

\begin{abstract}
We provide further evidence that the 
canonical quantization of cosmological models eliminates 
the classical Big Bang singularity, 
using the 
{\it DeBroglie-Bohm} interpretation of quantum mechanics.
The usual criterion for 
absence of the Big Bang singularity in Friedmann-Robertson-Walker 
(FRW) quantum cosmological models is the non-vanishing of the 
expectation value of the scale factor. 
 We 
compute the `local expectation value' of the Ricci and 
Kretschmann scalars, for some quantum FRW models. We show 
that they are finite for all time. 
Since these scalars are elements of general scalar polynomials in 
the metric and the Riemann tensor, this result indicates that, for 
the quantum models treated here, the `local expectation value' of 
these general scalar polynomials should be finite everywhere. 
Therefore, according to the classification introduced in Refs. 
\cite{ellis, tipler}, we have further evidence that the quantization 
of the models treated here eliminates the classical Big Bang singularity.


\end{abstract}

\maketitle

The presence of singularities in cosmological models in general
relativity is an old issue \cite{friedmann}. In a series of articles,
Hawking, Penrose and Geroch showed that, if certain very general
conditions are satisfied, singularities are always present in
cosmological models based on general relativity \cite{hawking}. One of
such singularities is the Big Bang singularity, which is believed to
represent the very beginning of the Universe. Here, one has a
fundamental problem because, if the beginning of the Universe is a
singular event of general relativity, that theory cannot describe it.
In order to overcome that fundamental problem, many authors proposed
the quantization of gravity. Quantum cosmology was the first of such
attempts and, since the first model, has showed good signs toward the
solution of the above mentioned problem \cite{dewitt}. Since then,
many important works have been done by computing the wave function of
the universe ($\Psi$) in different minisuperspace models. Some authors
find $\Psi$ by solving the Wheeler-DeWitt equation \cite{rubakov,
gotay, moss, vilenkin, lemos, alvarenga, germano, germano1, gil} and
others by using the path integral approach \cite{hawking1, fujiwara,
gil1}. As a common result, in most of them the problem of the initial
singularity was claimed to be solved. The main argument used to
support those claims depends on the quantum mechanical interpretation
used in each particular work.

The two interpretations most frequently used in quantum cosmology are
the {\it Many Worlds} one \cite{everett} and the {\it DeBroglie-Bohm}
one \cite{bohm}, \cite{holland}. As in the usual {\it Copenhaguen}
interpretation of quantum mechanics, in the {\it Many Worlds}
interpretation one cannot talk about trajectories of the canonical
variables, but only about mean values of those variables. On the other
hand, in the {\it DeBroglie-Bohm} interpretation the trajectories of
the canonical variables are meaningful and can, in principle, be
computed by solving a system of differential equations involving
derivatives of the wave function phase. In the minisuperspace models
\cite{misner} treated using the {\it Many Worlds} interpretation, the
common argument used to justify the absence of a Big Bang singularity
is the fact that the mean value of the scale factor ($a$), as a
function of a chosen time, never vanishes \cite{gotay, lemos, germano,
germano1, gil}. In the models investigated according to the {\it
DeBroglie-Bohm} interpretation, the argument was that the scale factor
Bohmian trajectories $a(t)$ as a function of a chosen time never go
through $a=0$ \cite{germano, germano1, acacio, acacio0}. That result
is supported by the fact that the quantum potential present in the
dynamical equation of $a$, for those models, is repulsive for $a$ near
to zero \cite{germano,germano1,acacio}. In the present work, we shall
restrict our attention to the {\it DeBroglie-Bohm} interpretation of
quantum mechanics.

Quantum cosmology was the first attempt in order to remove the Big
Bang singularity by quantizing the gravitational theory \cite{dewitt}.
The {\it DeBroglie-Bohm} interpretation of quantum mechanics
\cite{bohm}, \cite{holland}, is frequently used in quantum cosmology.
In the minisuperspace models treated using the {\it DeBroglie-Bohm}
interpretation the common argument used to justify the absence of a
Big Bang singularity is the fact that the scale factor Bohmian
trajectories $a(t)$, as a function of a chosen time,\textit{} never go
through $a=0$ \cite{germano, germano1, acacio, acacio0}. 

In order to derive the physical content of any operator $\hat{A}( x,
p_x)$, in the {\it DeBroglie-Bohm} interpretation of quantum
mechanics, one must compute the so-called `local expectation value' of
that operator defined as \cite{holland}

\begin{equation}
\label{1}
A(x,t) = Re\left( {\Psi^*(x,t)(\hat{A}\Psi)(x,t)\over
\Psi^*(x,t)\Psi(x,t)}\right)\, ,
\end{equation}
where $(\hat{A}\Psi)(x,t) = \int \hat{A}(x,x')\Psi (x',t)d^3x'$.
If we apply this definition in quantum cosmology, it is easy to see
that the `local expectation value' of the scale factor operator
is the real, time dependent, scale factor function. On the other hand,
the `local expectation value' of more complicated operators constructed
out of the scale factor and its canonically conjugate momentum will be
much more difficult to compute and in general will require a specific
factor ordering prescription.

%Based on an important singularities classification scheme proposed
%by G. F. R. Ellis and B. G. Schmidt \cite{ellis}, 
The Big Bang
singularities that occur in the classical Friedmann-Robertson-Walker (FRW)
models are said to be `scalar polynomial singularities' \cite{ellis}. 
A `scalar polynomial singularity' is the end point of at least one
curve on which a scalar polynomial in the metric and the Riemann
tensor becomes infinite \cite{tipler}. In the present
work, we wish to give further evidence, besides the usual one,
that the canonical quantization of FRW models removes the initial
Big Bang singularities of those models. We shall compute, for all
FRW models considered, the `local expectation value' of the Ricci
($\equiv g_{\alpha \beta} g_{\gamma \delta}
R^{\delta \alpha \gamma \beta}$) and the Kretschmann scalars
($\equiv R_{\alpha \beta \gamma \delta}
R^{\alpha \beta \gamma \delta}$). As we shall see, they are finite
for all time. Therefore, since these scalars are
components of general scalar polynomials in the metric and the
Riemann tensor, that result indicates that the
`local expectation value' of a general scalar polynomial in the
metric and the Riemann tensor should be free from singularities in
those models.

In the present work, we shall consider FRW cosmological models
coupled to a radiative perfect fluid, treated by means of the 
variational formalism developed by Schutz \cite{schutz}. Our main 
motivation to choose the matter content of the model as radiation 
is because we would like to describe the very early Universe, the
so called 'radiation dominated era'. At that time, 
the quantum effects were more important. The models are described 
by the Hamiltonian \cite{germano1}

\begin{equation}
{\cal H} = \frac{p_{a}^2}{24} + 6ka^2 - p_{T},
\label{2}
\end{equation}

\noindent
where $p_{a}$ and $p_{T}$ are, respectively, the momenta
canonically conjugate to the scale factor ($a$) and the
radiation variable ($T$). The parameter $k$ is related to the
spatial curvature of the model and may assume the values $+1$
(positive curvature), $-1$ (negative curvature) and zero (no
curvature). We employ the natural  system of units, defined by
$\hbar=c=16\pi G=1$.

The quantization of those models follows the Dirac formalism
for quantizing constrained systems. The application
of this formalism for the present models result in the following
Wheeler-DeWitt equation for the wave function $\Psi(a , T )$
\cite{germano1},

First we
introduce a wave function which is a function of the canonical
variables $a$ and $T$,

\begin{equation}
\label{3}
\Psi\, =\, \Psi(a , T )\, .
\end{equation}

\noindent
Then, we impose the appropriate commutators between the operators $a$
and $T$ and their conjugate momenta $p_a$ and $p_T$.
Working in the Schr\"{o}dinger picture, the operators $a$ and $T$
are simply multiplication operators, while their conjugate momenta are
represented by the differential operators
\begin{equation}
p_{a}\rightarrow -i\frac{\partial}{\partial a}\hspace{0.2cm},\hspace{0.2cm}
\hspace{0.2cm}p_{T}\rightarrow -i\frac{\partial}{\partial T}\hspace{0.2cm}.
\label{4}
\end{equation}

Finally, we demand that the operator corresponding to $\mathcal{H}$
annihilate the wave function $\Psi$, which leads to the Wheeler-DeWitt
equation. From Eq. (\ref{2}), this formalism leads to

\begin{equation}
\frac{\partial^2 \Psi}{\partial a^2}-144ka^2\Psi+
24i\frac{\partial \Psi}{\partial \tau}=0,
\label{5}
\end{equation}
where $T=-\tau$. Several solutions to equation
(\ref{5}) are known.

Consider first the case $k=0$. Then the solution to Eq.
(\ref{5}) is given by \cite{lemos},

\begin{equation}
\Psi (a, \tau) = \left(\frac{4\sigma}
{\pi}\right)^{1/4}\sqrt{\frac{1}{
(2\sigma\tau-i)(2\tau+i)}}
\times \exp\left\{
\frac{6i}{\tau}\left[
1+\frac{i}{2\sigma\tau-i}\right]a^2\right\}.
\label{6}
\end{equation}

In the cases $k=\pm 1$, the solution of Eq. (\ref{5}) may be
written in the following form \cite{germano1}:

\begin{eqnarray}
\Psi (a, \tau) & = & \left(\frac{4\sigma}
{\pi}\right)^{1/4}\sqrt{\frac{k}{\cos^2(\sqrt{k}\tau)
[2\sigma\tan(\sqrt{k}\tau)-i\sqrt{k}][2\tan(\sqrt{k}\tau)+
i\sqrt{k}]}}\nonumber \\
& \times & \exp\left\{
\frac{6i\sqrt{k}}{\tan(\sqrt{k}\tau)}\left[
1+\frac{i\sqrt{k}}{\cos^2(\sqrt{k}\tau)[2\sigma\tan(\sqrt{k}\tau)
-i\sqrt{k}]}\right]a^2\right\}.
\label{7}
\end{eqnarray}

In order to use the {\it DeBroglie-Bohm} interpretation we must
rewrite $\Psi$ in the polar form

\begin{equation}
\Psi(a,\tau) = \left({4\sigma\over\pi}\right)^{1/4}\Theta\exp{(iS)}\,
\label{8}
\end{equation}

Consider first the case $k=0$. Then the solution to Eq.
(\ref{5}) is given in polar form by \cite{lemos},

\begin{equation}
\label{9}
\Theta\, =\, g_0 (\tau)\times \exp{\left({-12\sigma a^2\over 1 +
4\sigma^2\tau^2}\right)}\qquad \mbox{and}
\end{equation}
\begin{equation}
\label{10}
S\, =\, f_0 (\tau) + \left({24\sigma^2\tau\over 1 +
4\sigma^2\tau^2}\right) a^2\, .
\end{equation}

For the cases $k=\pm 1$ one has \cite{germano1}

\begin{equation}
\label{11}
\Theta\, =\, g_k (\tau)\times \exp{\left({-12\sigma k a^2\over
k\cos^2{(\sqrt{k}\tau)} + 4\sigma^2\sin^2{(\sqrt{k}\tau)}}\right)}
\qquad \mbox{and}
\end{equation}
\begin{equation}
\label{12}
S\, =\, f_k (\tau) + {12\sqrt{k}\over 2\tan{(\sqrt{k}\tau)}}
\left[ 1 - {k\over k\cos^2{(\sqrt{k}\tau)} +
4\sigma^2\sin^2{(\sqrt{k}\tau)}}\right] a^2\, .
\end{equation}

Following the {\it DeBroglie-Bohm} interpretation and the
fact that $p_a = 12 \dot{a}$, where the dot means
differentiation with respect to $\tau$, we may compute the
scale factor Bohmian trajectory from \cite{acacio}

\begin{equation}
\label{13}
p_a = {\partial S\over \partial a}\, .
\end{equation}
%Also, we may compute the quantum potential $Q$
%from \cite{acacio}

%\begin{equation}
%\label{14}
%Q = - {1\over 24 \Theta} {\partial^2 \Theta\over \partial a^2}\, .
%\end{equation}

For the case $k=0$,  from Eqs. (\ref{9}) and
(\ref{10}) we have \cite{germano1, acacio0}

\begin{equation}
\label{15}
a(\tau) = a_0(1+4\sigma^2 \tau^2)^{1/2},
\end{equation}
where $a_0$ is an integration constant and
\begin{equation}
\label{16}
Q = {\sigma\over 1+4\sigma^2\tau^2} -
{24\sigma^2 a^2\over (1+4\sigma^2\tau^2)^2}\, .
\end{equation}
For the cases $k=\pm 1$,  Eqs. (\ref{11}) and
(\ref{12}) yield \cite{germano1, acacio0}

\begin{equation}
\label{17}
a(\tau) = a_k(k\cos^2{(\sqrt{k}\tau )} +
4\sigma^2\sin^2{(\sqrt{k}\tau)})^{1/2},
\end{equation}
where $a_k$ is an arbitrary integration constant, 
associated to a given $k$, and
\begin{equation}
\label{18}
Q = {\sigma k\over k\cos^2{(\sqrt{k}\tau )} +
4\sigma^2\sin^2{(\sqrt{k}\tau)}} - {24\sigma^2 k^2 a^2
\over (k\cos^2{(\sqrt{k}\tau )} +
4\sigma^2\sin^2{(\sqrt{k}\tau)})^2}\, .
\end{equation}

In all these cases it is clear that the scale factor Bohmian
trajectories never reach $a=0$. Therefore, one may conclude
that these models are free from the Big Bang singularity.

Here, we would like to give further evidence, besides the
non-vanishing of the scale factor, that those models are free
from the Big Bang singularity at the quantum level. In
order to do that, we shall compute the `local expectation
value' of the Ricci and the Kretschmann scalars.
As it will be seen, they remain finite for all time.
These scalars are components of general scalar polynomials in 
the metric and the Riemann tensor. Therefore, we have an 
additional indication that the `local expectation value' of 
these general scalar polynomials should be finite everywhere.

For the present models, the expressions of the Ricci ($R$) and
the
Kretschmann ($K$) scalars are given, respectively, by

\begin{equation}
R = g^{ac}g^{bd}R_{abcd} = \frac{\dot{p}_a}{2a^3} +
{6k\over a^2}\qquad \mbox{and}
\label{19}
\end{equation}
\begin{equation}
K = R^{abcd}R_{abcd} =
\frac{\dot{p}^2_a}{12a^6} - {\dot{p}_a p^2_a\over 72a^7} +
{p^4_a\over 864 a^8} + {k p^2_a\over 6 a^6} +
{12 k^2\over a^4},
\label{20}
\end{equation}
where we used  $p_a = 12\dot{a}$ in order to write $R$ and $K$ 
in terms of ${p}_a$, the momentum canonically conjugate to ${a}$.

It is important to notice, before we proceed, that $R$ and
$K$ given by Eqs. (\ref{19}) and (\ref{20}) are to be promoted
to quantum
operators. Since, at the quantum level, ${p}_a$ and ${a}$
do not commute, we shall have to introduce a specific factor
ordering in order to correctly describe the terms involving
products of powers of $a$ and $p_a$. Here,
we shall use a symmetrization procedure known as the Weyl
ordering \cite{lee}. In order to obtain
the Weyl-ordered expression of a product ($a^n p^m_a$),
one first randomly orders the $a's$ and $p_a's$, with each
different ordering counted once, then divides the result
by the number of terms present in the final expression
\cite{lee}.

In Eqs. (\ref{19}) and (\ref{20}) we notice the presence
of the time rate of change of the momentum $p_a$. Quantum
mechanically, it is an operator and in the {\it DeBroglie-Bohm}
interpretation it has the following value \cite{holland},

\begin{equation}
\label{21}
\dot{p}_a = - \nabla (V + Q)\,
\end{equation}
Where $V$ is the classical potential present in the Hamiltonian
(\ref{2}) and $Q$ is the quantum potential. From Eq. (\ref{2})
$V$ is given by
\begin{equation}
\label{22}
V = 6 k a^2\, .
\end{equation}
Using expressions (\ref{16}) and (\ref{18}) for $Q$, and
Eq. (\ref{22}) for $V$, we may compute the time rate of change of the
momentum $p_a$.

For $k=0$ we obtain

\begin{equation}
\label{23}
\dot{p}_a = {48\sigma^2 a_0^4\over a^3}\, ,
\end{equation}

while for $k=\pm 1$ we find

\begin{equation}
\label{24}
\dot{p}_a = -12ka + {48\sigma^2 a_k^4\over a^3}\, .
\end{equation}
Introducing the above values of $\dot{p}_a$ in the expressions
for the Ricci and Kretschmann scalars we obtain new forms
which depend only on the operators $a$ and $p_a$.

For $k=0$, introducing $\dot{p}_a$  given by  Eq. (\ref{23}) in Eqs. (\ref{19})
and (\ref{20}), we get

\begin{equation}
\label{25}
R = {24\sigma^2\over a_0^2}{1\over (1+4\sigma^2\tau^2)^3}\, ,
\end{equation}
\begin{equation}
\label{26}
K = {1\over 864}{p_a^4\over a^8} - {2a_0^4\sigma^2\over 3}{p_a^2\over a^{10}}
+ 192\sigma^4a_0^8 {1\over a^{12}}\, .
\end{equation}

For $k=\pm 1$, inserting $\dot{p}_a$ from Eq. (\ref{24}) into Eqs. (\ref{19})
and (\ref{20}), we obtain

\begin{equation}
\label{27}
R = {24\sigma^2\over a_k^2}{1\over (k\cos^2{(\sqrt{k}\tau )} +
4\sigma^2\sin^2{(\sqrt{k}\tau)})^3}\, ,
\end{equation}
\begin{equation}
\label{28}
K = {1\over 864}{p_a^4\over a^8} - {2a_k^4\sigma^2\over 3}{p_a^2\over a^{10}}
+ {k\over 3}{p_a^2\over a^6} + 192\sigma^4a_k^8 {1\over a^{12}} -
96\sigma^2 k^3 a_k^4{1\over a^8} + 24{1\over a^4}\, .
\end{equation}

\begin{figure}[htb]
\begin{minipage}[t]{0.49\textwidth}
%\vspace{3.0cm}
\includegraphics[width=5cm,height=6cm,angle=0]{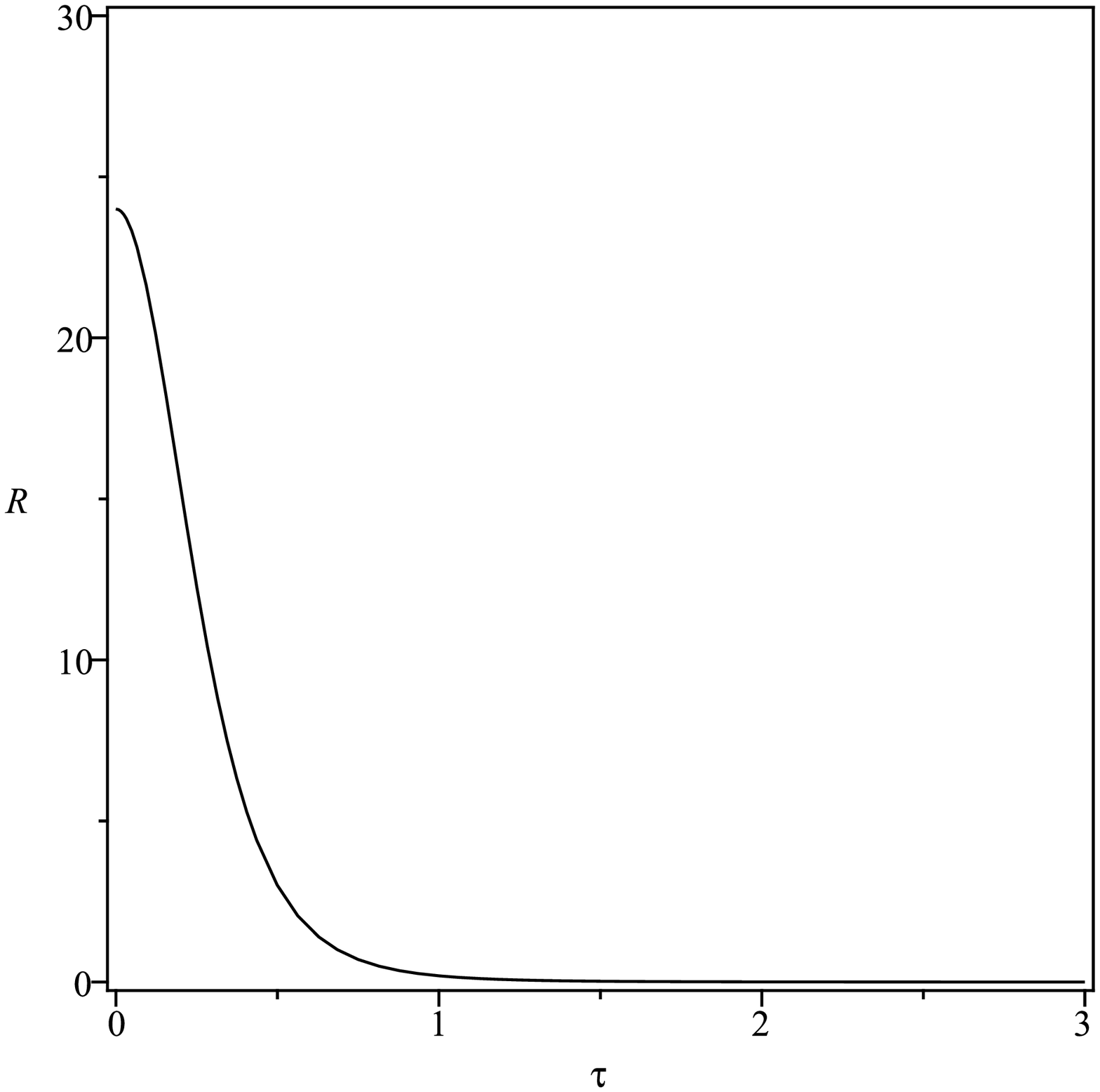}
\caption{The $R$ `local expectation value' for $k=0$,
$\sigma=1$ and $a_{0}=1$. It is always regular and goes 
to zero for large $\tau$. For this case $R$ evaluated 
over the classical scale factor is identically zero.}
\label{f1}
\end{minipage}
\hspace{0cm}
\begin{minipage}[t]{0.49\textwidth}
%\vspace{3.0cm}
\includegraphics[width=5cm,height=6cm,angle=0]{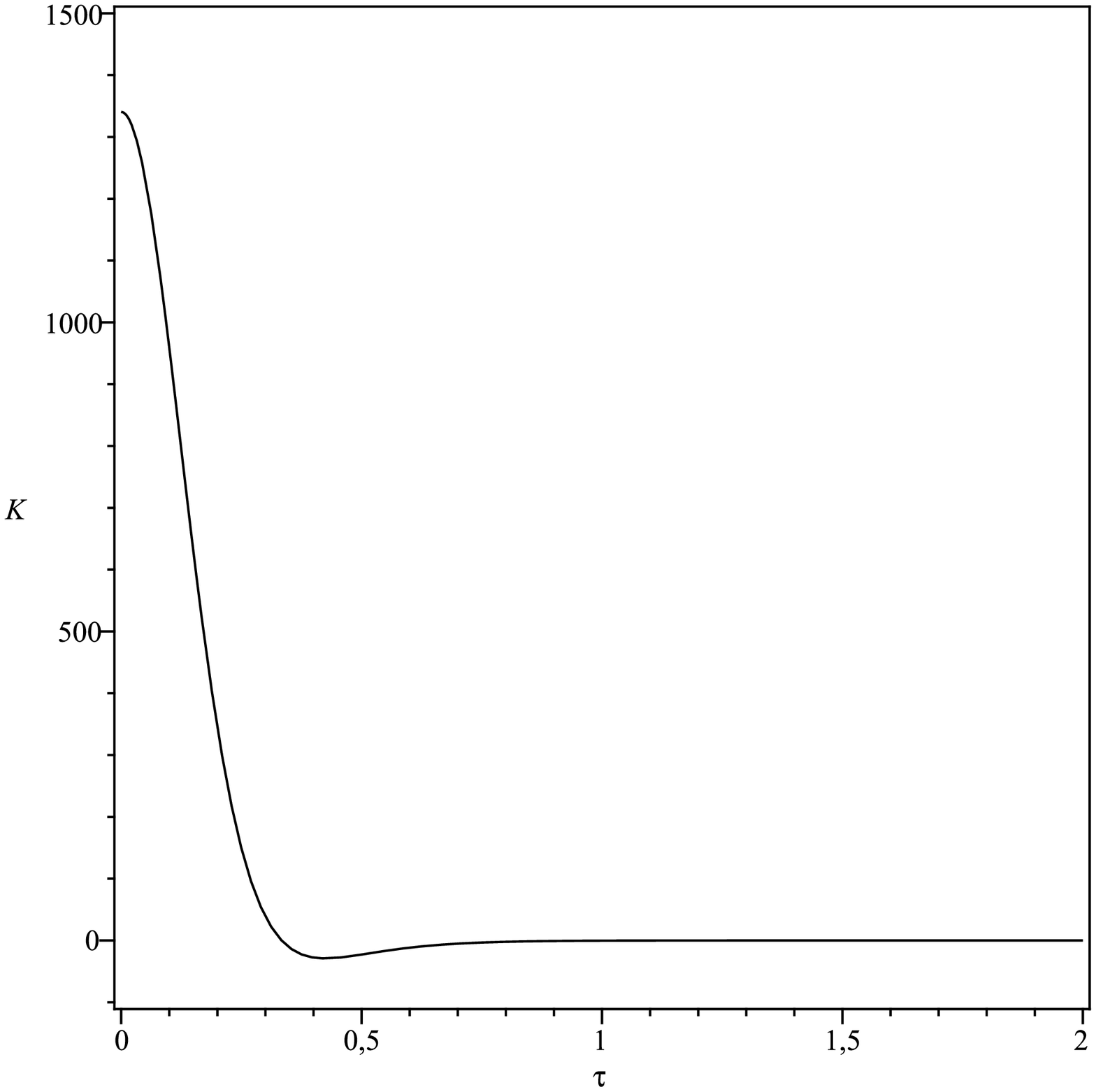}
\caption{The $K$ `local expectation value' for $k=0$,
$\sigma=1$ and $a_{0}=1$. It is always regular and goes 
to zero for large $\tau$. For this case $K$ evaluated 
over the classical scale factor is identically zero.}
\label{f2}
\end{minipage}
\end{figure}

\begin{figure}[htb]
\begin{minipage}[t]{0.49\textwidth}
%\vspace{3.0cm}
\includegraphics[width=5cm,height=6cm,angle=0]{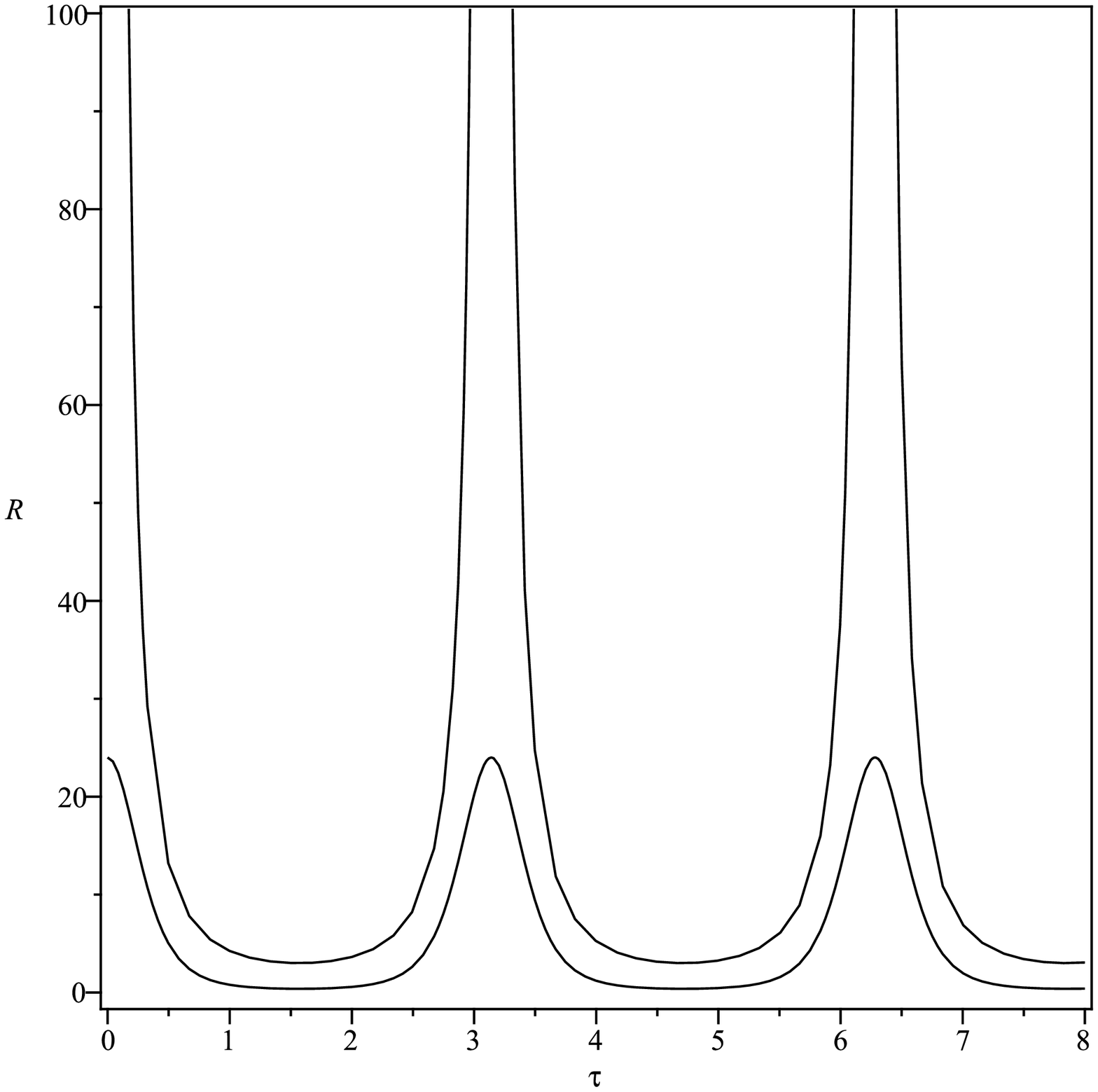}
\caption{The lower curve represents the $R$ `local 
expectation value' for $k=1$, $\sigma=1$ and $a_{1}=1$. 
It is periodic and always regular. The upper curve 
represents $R$ evaluated over the classical scale factor.}
\label{f3}
\end{minipage}
\hspace{0cm}
\begin{minipage}[t]{0.49\textwidth}
%\vspace{3.0cm}
\includegraphics[width=5cm,height=6cm,angle=0]{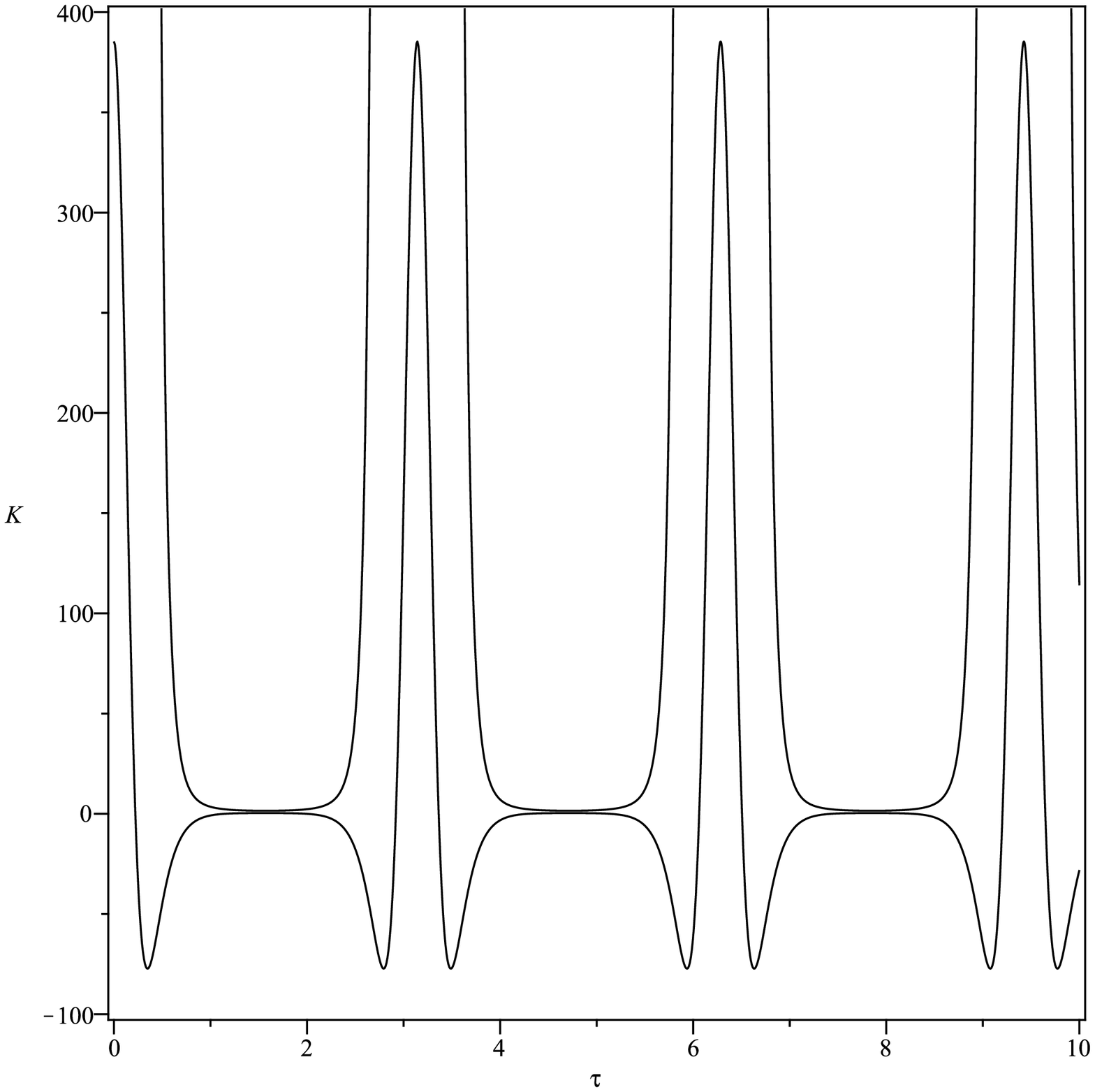}
\caption{The lower curve represents the $K$ `local 
expectation value' for $k=1$, $\sigma=1$ and $a_{1}=1$. 
It is periodic and always regular. The upper curve 
represents $K$ evaluated over the classical scale factor.}
\label{f4}
\end{minipage}
\end{figure}

\begin{figure}[htb]
\begin{minipage}[t]{0.49\textwidth}
%\vspace{3.0cm}
\includegraphics[width=5cm,height=6cm,angle=0]{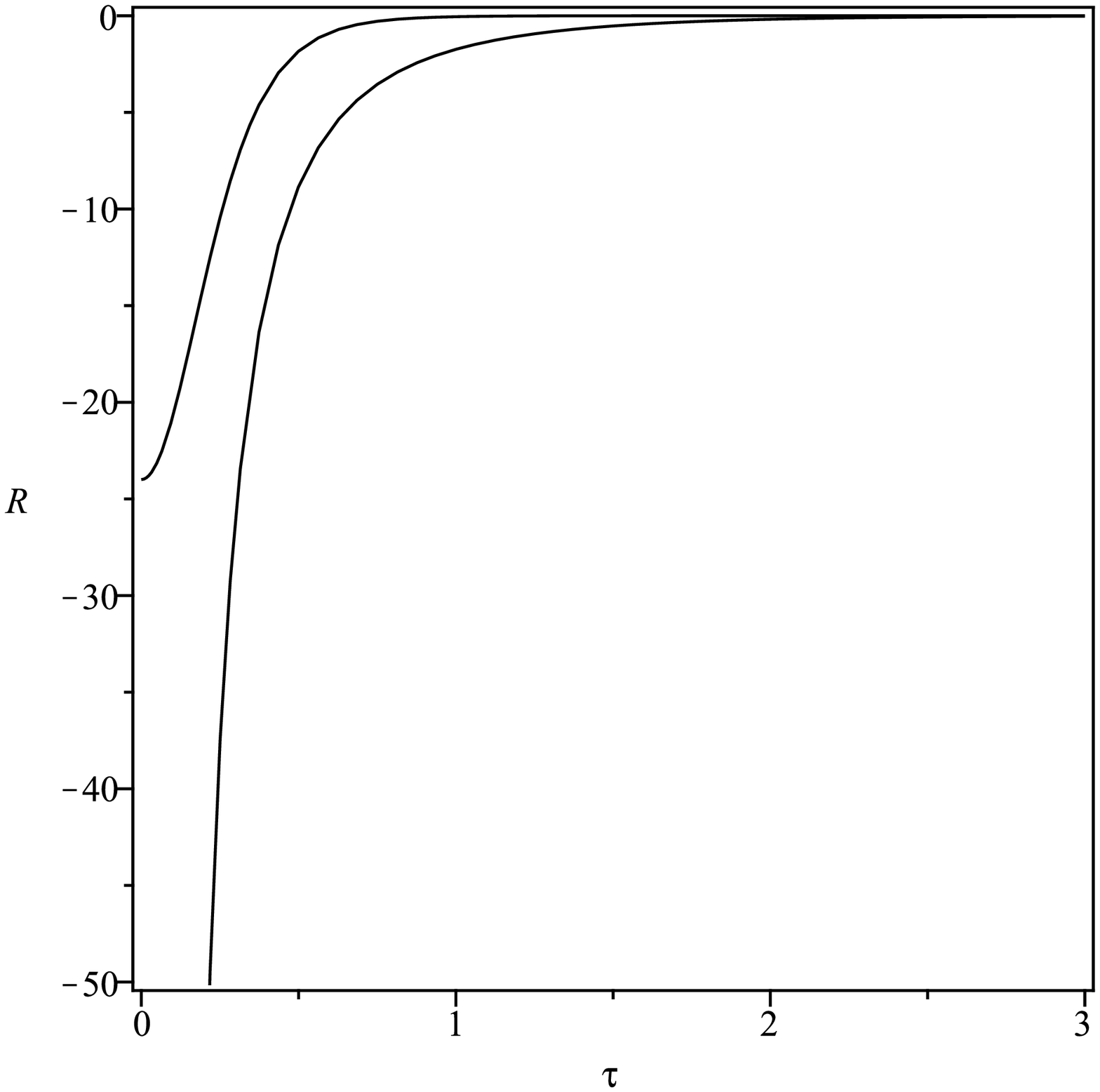}
\caption{The upper curve represents the $R$ `local 
expectation value' for $k=-1$, $\sigma=1$ and $a_{-1}=1$. 
It is always regular. The lower curve represents $R$ 
evaluated over the classical scale factor. Both 
curves go to zero for large $\tau$.}
\label{f5}
\end{minipage}
\hspace{0cm}
\begin{minipage}[t]{0.49\textwidth}
%\vspace{3.0cm}
\includegraphics[width=5cm,height=6cm,angle=0]{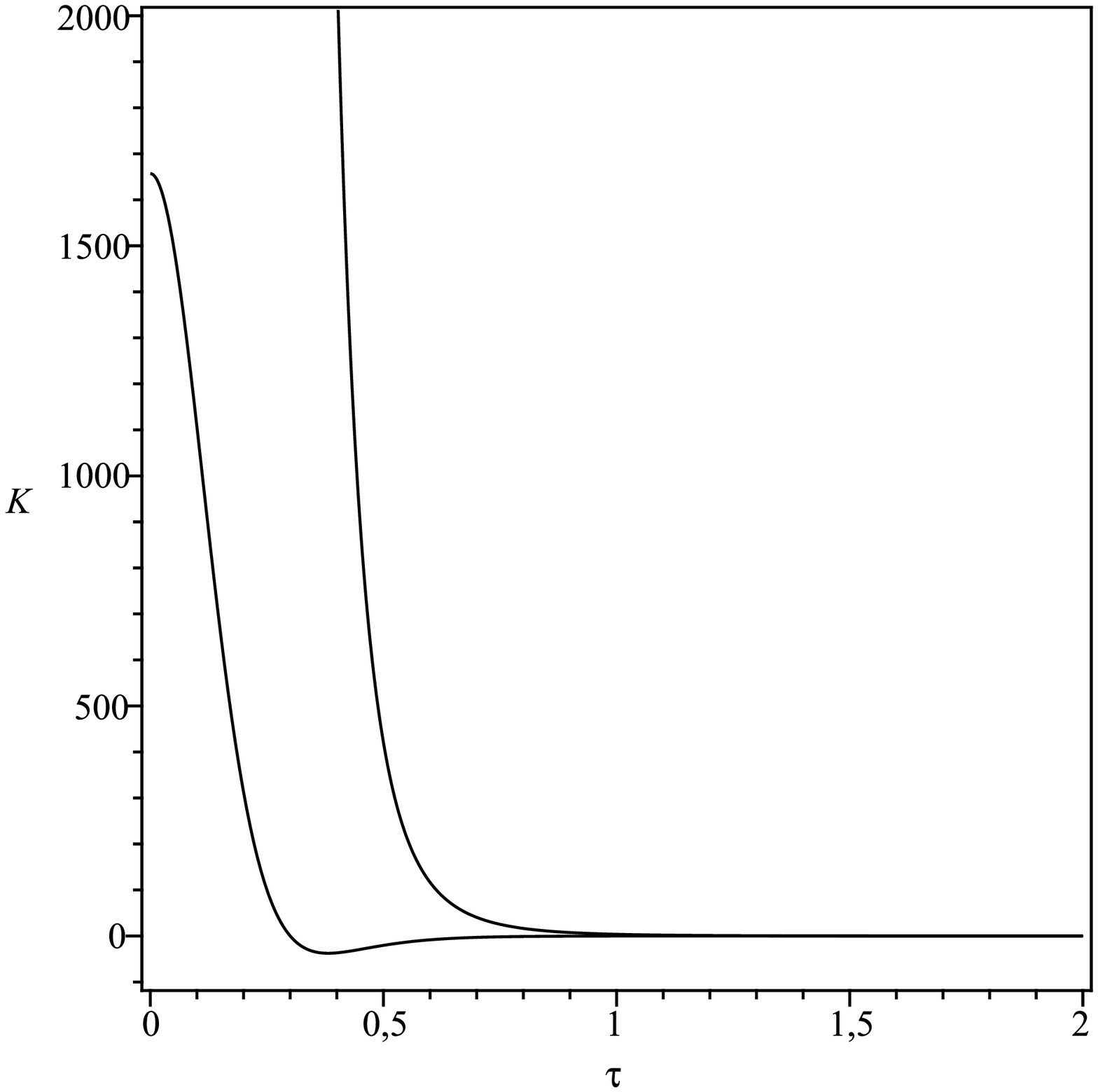}
\caption{The lower curve represents the $K$ `local 
expectation value' for $k=-1$, $\sigma=1$ and $a_{-}=1$. 
It is always regular. The upper curve represents $K$ 
evaluated over the classical scale factor. Both 
curves go to zero for large $\tau$.}
\label{f6}
\end{minipage}
\end{figure}

The expressions for $R$ both in the cases $k=0$ and $k=\pm 1$ depend only on
the operator $a$. Therefore, the $R$ `local expectation value', for both
cases, is given by the same expression of the operator $R$ with the
operator $a$ replaced by the real, time dependent scalar factor function.
It is clear from the above expressions for $R$ that it is  regular at $\tau = 0$. 
Thus, $R$ is regular for all values of $k$ at the beginning moment of the 
universes described by the corresponding models. In order to compare the 
behavior of the $R$ `local expectation value', as a function of $\tau$, 
with the classical expression of $R$, we produced Figs. (\ref{f1}), 
(\ref{f3}) and (\ref{f5}), one for each value of $k$. The classical scalar
factor was derived with initial conditions compatible with those of the 
scale factor Bohmian trajectories. One may easily see from those figures that, 
for the cases where the spatial sections are open, both quantities coincide for 
large $\tau$.

In order to derive the physical content of the operator $K$, in the {\it
DeBroglie-Bohm} interpretation of quantum mechanics, we must compute its
`local expectation value' Eq. (\ref{1}). Due to the presence of the terms

\begin{equation}
\label{29}
{p_a^4\over a^8}, \qquad {p_a^2\over a^{10}},\qquad {p_a^2\over a^6}
\end{equation}
in the expressions of $K$, Eqs. (\ref{26}) and (\ref{28}), we choose to
use the Weyl ordering \cite{lee}. 

In order to reduce the Weyl-ordered expressions of
each one of the above products of $a's$ and $p_a's$, we use the commutation
relation between the operators $a$ and $p_a$. With its aid, we write all
terms, in each of the Weyl-ordered expressions, such that $p_a$ or a power
of $p_a$ must appear to the right of $a$ or a power of $a$. This procedure
simplifies very much the Weyl-ordered expressions because most of the
terms combine with each other. 
With the aid of the following commutators,

\begin{eqnarray}
\label{30}
\left[ a^n , p_a^2 \right] & = & 2in a^{n-1} p_a + n(n-1) a^{n-2}\nonumber\\
\left[ a^n , p_a^4 \right] & = & 4in a^{n-1} p_a^3 + 6n(n-1) a^{n-2} p_a^2
- 4in(n-1)(n-2) a^{n-3} p_a\nonumber\\
& - & n(n-1)(n-2)(n-3) a^{n-4},
\end{eqnarray}
where $n$ is a positive or negative integer, 
we obtain the following Weyl-ordered expressions for each one of the
products of $a's$ and $p_a's$ in Eq. (\ref{29}):

\begin{eqnarray}
\label{31}
\left({p_a^4\over a^8}\right)_W & = & {1\over a^8} p_a^4 + 16i{1\over a^9} p_a^3
- 116{1\over a^{10}} p_a^2 - 440i{1\over a^{11}} p_a + 10831{1\over 15 a^{12}},
\nonumber\\
\left({p_a^2\over a^{10}}\right)_W & = & {1\over a^{10}} p_a^2 +
10i{1\over a^{11}} p_a - 175{1\over 6 a^{12}},\nonumber\\
\left({p_a^2\over a^6}\right)_W & = & {1\over a^6} p_a^2 + 6i{1\over a^7} p_a
- 23{1\over 2 a^8}.
\end{eqnarray}

Now, we substitute
$p_a$ given by $-i\partial/\partial a$ into these Weyl-ordered expressions and 
compute their `local expectation value' Eq. (\ref{1}), using the wave function
(\ref{8}). 
It is important to remember that
the wave function $\Psi$ must be written in the polar form  (\ref{8}). 
Then, we obtain the expressions for the `local expectation values' of
each operator in Eq. (\ref{31}) as functions of $\Theta (a,\tau)$ and 
derivatives of $\Theta (a,\tau)$ and $S(a,\tau)$ with respect to $a$.

\begin{eqnarray}
\label{32}
\left\langle {p_a^4\over a^8} \right\rangle_L & = &
{(\partial S(a, \tau)/\partial a)^4\over a^8} - {3(\partial^2
S(a, \tau)/\partial a^2)^2 \over a^8} - {116(\partial
S(a, \tau)/\partial a)^2\over a^{10}}\nonumber\\
& - & {12(\partial \Theta(a, \tau)/\partial a)(\partial^2 S(a, \tau)/\partial a^2)
(\partial S(a, \tau)/\partial a)\over a^8 \Theta(a, \tau)}\nonumber\\
& + & {48(\partial^2 S(a, \tau)/
\partial a^2)(\partial S(a, \tau)/\partial a)\over a^9}\nonumber\\
& - & {6(\partial^2 \Theta(a, \tau)/\partial a^2)(\partial S(a, \tau)/\partial a)^2
\over a^8 \Theta(a, \tau)} - {16(\partial^3 \Theta(a, \tau)/\partial a^3)\over a^9 \Theta(a, \tau)}\nonumber\\
& + & {116(\partial^2 \Theta(a, \tau)/\partial a^2)\over a^{10} \Theta(a, \tau)} +
{\partial^4 \Theta(a, \tau)/\partial a^4)\over a^8 \Theta(a, \tau)} -
{440(\partial \Theta(a, \tau)/ \partial a)\over a^{11} \Theta(a, \tau)}\nonumber\\
& + & {48(\partial \Theta(a, \tau)/\partial a)(\partial S(a, \tau)/\partial a)^2\over
a^9 \Theta(a, \tau)} - {4(\partial^3 S(a, \tau)/\partial a^3)(\partial S(a, \tau)\partial a)\over a^8}\nonumber\\
& + & {10831\over 15 a^{12}},
\end{eqnarray}
\begin{equation}
\label{33}
\left\langle {p_a^2\over a^{10}} \right\rangle_L = - {175\over 6 a^{12}}
+ {(\partial S(a, \tau)/ \partial a)^2\over a^{10}} +
{10(\partial \Theta(a, \tau)/ \partial a)\over a^{11} \Theta(a, \tau)}
- {(\partial^2 \Theta(a, \tau)/ \partial a^2)\over a^{10} \Theta(a, \tau)},
\end{equation}
\begin{equation}
\label{34}
\left\langle {p_a^2\over a^6} \right\rangle_L = - {23\over 2 a^8} +
{6(\partial \Theta(a, \tau)/ \partial a)\over a^7 \Theta(a, \tau)} -
{(\partial^2 \Theta(a, \tau)/ \partial a^2)\over a^6 \Theta(a, \tau)} +
{(\partial S(a, \tau)/ \partial a)^2\over a^6}.
\end{equation}

Finally, we compute the $K$ `local expectation value' for both
cases of $k=0$ Eq. (\ref{26}) and $k=\pm 1$ Eq. (\ref{28}). 
Then, in order to compute the $K$ `local expectation value', 
After that, we 
introduce the values of $\Theta(a, \tau )$ and $S(a, \tau )$
for each $k$, Eqs. (\ref{9}-\ref{12}) in the `local expectation values'
of $(p_a^4/ a^8)_W$, $(p_a^2/ a^{10})_W$ and $(p_a^2/ a^6)_W$.

Finally, we combine them following
Eq. (\ref{26}) for $k=0$ and Eq. (\ref{28}) for $k=\pm 1$.

For $k=0$ we find

\begin{equation}
\label{35}
\left< K \right>_L = 
{\frac {
C_8^0 a_0^8
+ C_6^0 a_0^6
+ C_4^0 a_0^4
+ C_2^0 a_0^2
+ C_0^0} 
{a_0^{12} \left( 1 + 4\sigma^2\tau^2 \right)^6}},
\end{equation}
where
\begin{eqnarray}
\label{36}
C_8^0 &=&  6144\sigma^8\tau^4 - 10752\sigma^6\tau^2 + 960\sigma^4,\nonumber \\
C_6^0 &=& - 1920\sigma^5\tau^2 + 304\sigma^3,\nonumber \\
C_4^0 &=& - 56\sigma^4\tau^2 + {601\over 9}\,\sigma^2,\nonumber  \\
C_2^0 &=& 9\sigma,\nonumber \\
C_0^0 &=& {10831\over 12960}.
\end{eqnarray}
For $k = 1$ we get

\begin{equation}
\label{37}
\left< K \right>_L =
{\frac {
C_8^1 a_1^8
+ C_6^1 a_1^6
+ C_4^1 a_1^4
+ C_2^1 a_1^2
+ C_0^1}
{a_1^{12} \left( \cos^2{\tau } +
4\sigma^2\sin^2{\tau} \right)^6}},
\end{equation}
where
\begin{eqnarray}
\label{38}
C_8^1 &=& 24 ( 256\sigma^8 + 256\sigma^6 - 160\sigma^4 + 16\sigma^2 + 1 )\cos^4{\tau}\nonumber \\
&+& 96\sigma^2 ( -128\sigma^6 - 16\sigma^4 + 40\sigma^2 - 7 )\cos^2{\tau}\nonumber \\
&+& 192\sigma^4 (32\sigma^4 - 24\sigma^2 + 5 ),\nonumber \\
C_6^1 &=& 80\sigma ( 16\sigma^4 - 8\sigma^2 + 1 )\cos^4{\tau}
+ 40\sigma ( -16\sigma^4 + 16\sigma^2 \nonumber \\
&-& 3 )\cos^2{\tau} - 16\sigma^3 ( 40\sigma^2 + 19 ),\nonumber \\
C_4^1 &=& (17/15)( -4\sigma^4 + 2\sigma^2 - (1/4))\cos^4{\tau} + (1/5)( (988/3)\sigma^4 \nonumber \\
&-& (34/3)\sigma^2 - (71/4))\cos^2{\tau} + (2/3)\sigma^2 ( - 92\sigma^2 + (757/15)),\nonumber  \\
C_2^1 &=& (27/10)\sigma,\nonumber \\
C_0^1 &=& (10831/12960).
\end{eqnarray}
Finally, for $k = - 1$ we obtain

\begin{equation}
\label{39}
\left< K \right>_L =
{\frac {
C_8^{-1} a_{-1}^8
+ C_6^{-1} a_{-1}^6
+ C_4^{-1} a_{-1}^4
+ C_2^{-1} a_{-1}^2
+ C_0^{-1}}
{a_{-1}^{12} \left( \cosh^2{\tau } + 
4\sigma^2\sinh^2{\tau } \right)^6}},
\end{equation}
where
\begin{eqnarray}
\label{40}
C_8^{-1} &=& 24 ( 256\sigma^8 - 256\sigma^6 - 160\sigma^4 - 16\sigma^2 + 1 )\cosh^4{\tau}\nonumber \\
&+& 96\sigma^2 ( - 128\sigma^6 + 16\sigma^4 + 40\sigma^2 + 7 )\cosh^2{\tau}\nonumber \\
&+& 192\sigma^4 ( 32\sigma^4 + 24\sigma^2 + 5 ),\nonumber \\
C_6^{-1} &=& 80\sigma ( - 16\sigma^4 - 8\sigma^2 - 1 )\cosh^4{\tau} 
+ 40\sigma ( 16\sigma^4 + 16\sigma^2 \nonumber \\
&+& 3 )\cosh^2{\tau} + 16\sigma^3 ( 40\sigma^2 + 19 ),\nonumber \\
C_4^{-1} &=& (17/15)( 4\sigma^4 + 2\sigma^2 + (1/4))\cosh^4{\tau} + (1/5)( - (988/3)\sigma^4 \nonumber \\
&-& (34/3)\sigma^2 + (71/4))\cosh^2{\tau} + (2/3)\sigma^2 ( 92\sigma^2 + (757/15)),\nonumber  \\
C_2^{-1} &=& (27/10)\sigma,\nonumber \\
C_0^{-1} &=& (10831/12960).
\end{eqnarray}

It is clear from Eqs. (\ref{35}), (\ref{37}) and (\ref{39})
that the $K$ `local expectation value', for each model, is regular 
for all $\tau$. Including the limit  $\tau \to 0$, that is, at the 
beginning moments of the corresponding classical universes.
In order to compare the behavior of the $K$ `local expectation value', 
as a function of $\tau$, with the classical expression of $K$, we produced 
Figs. (\ref{f2}), (\ref{f4}) and (\ref{f6}), one for each value of $k$. 
The classical scalar factor was derived with initial conditions compatible 
with those of the scale factor Bohmian trajectories. One may easily see from 
those figures that, for the cases where the spatial sections are open, both 
quantities coincide for large $\tau$.
It is important to notice, that this result is independent of the
factor ordering used here. In fact, observing Eqs. (\ref{35}), 
(\ref{37}) and (\ref{39}), we conclude that they are regular mainly
because the scale factor Bohmian trajectories $a(t)$ never go through 
$a=0$. Therefore, from the 
operatorial expression of $K$ (\ref{28}), it is not difficult to see
that whatever factor ordering we decide to use the denominator of the 
$K$ `local expectation value' will be a polynomial in the scale factor. 
Then, if we take in account that it does not vanish for any $\tau$, the 
$K$ `local expectation value' will always be regular.
Now, since, $R$ and $K$ are elements of general scalar polynomials in 
the metric and the Riemann tensor, the above results indicate that, for 
the quantum models treated here, the `local expectation value' of 
these general scalar polynomials should be free of singularities. 
Therefore, according to the classification introduced in Refs. 
\cite{ellis,tipler}, we have further evidence that the quantization 
of the models treated here eliminates the classical Big Bang singularity.

We believe that the above result may be extended to FRW models with
matter contents described by other types of perfect fluids. This is
the case because, as we have mentioned above, the main reason for
the regularity of the `local expectation values' of $R$ and $K$ is that
the scale factor Bohmian trajectories $a(t)$, as a function of a chosen 
time, never go through $a=0$. Therefore, if we consider FRW models with
matter contents described by other types of perfect fluids and we obtain
Bohmian trajectories $a(t)$ that never go through $a=0$, very likely
the `local expectation values' of $R$ and $K$ of those models will
always be regular. In Reference \cite{germano}, the authors calculated
the scale factor Bohmian trajectory, as a function of the proper time, 
for a flat ($k=0$) FRW model with matter content described by a generic
perfect fluid. A generic perfect fluid is described by the equation of 
state $p=w\rho$, where $w$ is a constant. In the radiation case 
$w=1/3$. There, they found that the scale factor never go through 
$a=0$ whatever value $w$ assumes. Therefore, for that case we
are very confident to say that the `local expectation values' of $R$ and 
$K$ should always be regular. Unfortunately, we do not have such general 
result for models with $k=\pm 1$. On the other hand, if one believes that 
the quantization of those models will solve the singularity problem, they
will also have scale factor Bohmian trajectories that never go through $a=0$.
In this way, the `local expectation values' of $R$ and $K$ of those models 
should, also, always be regular.

{\bf Acknowledgements.} G. Oliveira-Neto, E. V. Corr\^{e}a Silva and  G. A. 
Mo\-ne\-rat (Researchers of CNPq, Brazil) thank CNPq and FAPERJ
for partial financial support. We thank the opportunity to use the 
Laboratory for Advanced Computation (LCA) of the Department of Mathematics 
and Computation, FAT/UERJ, where part of this work was prepared.

\end{document}